# How fast are the motions of tertiary-structure elements in proteins?

Gilad Haran[1]* and Hisham Mazal[1,#]

Department of Chemical and Biological Physics, Weizmann Institute of Science, Rehovot, Israel

*e-mail: gilad.haran@weizmann.ac.il

**Abstract**

Protein motions occur on multiple time and distance scales. Large-scale motions of protein tertiary-structure elements, i.e. domains, are particularly intriguing as they are essential for the catalytic activity of many enzymes and for the functional cycles of protein machines and motors. Theoretical estimates suggest that domain motions should be very fast, occurring on the nanosecond or microsecond time scales. Indeed, free-energy barriers for domain motions are likely to involve salt bridges, which can break in microseconds. Experimental methods that can directly probe domain motions on fast time scales have appeared only in recent years. This Perspective discusses briefly some of these techniques, including NMR and single-molecule fluorescence spectroscopies. We introduce a few recent studies that demonstrate ultrafast domain motions, and discuss their potential roles. Particularly surprising is the observation of tertiary-structure element dynamics that are much faster than the functional cycles in some protein machines. These swift motions can be rationalized on a case-by-case basis. For example, fast domain closure in multi-substrate enzymes may be utilized to optimize relative substrate orientation. Whether a large mismatch in time scales of conformational dynamics vs. functional cycles is a general design principle in proteins remains to be determined.

---

[#] Hisham Mazal's current address is: Max Planck Institute for the Science of Light, 91058 Erlangen, Germany

**Introduction**

Proteins are the key functional molecules in the living system, governing nearly all cellular functions and biochemical tasks. For example, proteins facilitate protein folding and prevent protein aggregation, they regulate membranal transport of ions and other chemicals, they participate in metabolic pathways and in many other cellular processes (1-4). Many proteins operate as machines, which can be quite broadly and somewhat loosely defined as devices that utilize an external energy source to perform a function. In order to coordinate their activities spatially and temporally, multiple protein machines have developed mechanisms for sophisticated regulation. Multiple internal motions in proteins are involved in such regulatory activities, and mapping them has become a major endeavor of modern biophysics, with far-reaching implications not only in basic science but also for the design of new protein functionalities and novel drugs.

The structure of proteins, even large ones, can nowadays be determined using one of an array of methods, from x-ray crystallography (5) through nuclear magnetic resonance spectroscopy (6) to cryo-electron microscopy (7). However, structural models provide only a static picture of a protein, and may miss the rich dynamics that can typically occur on multiple time and length scales (8, 9). Importantly, the conformational dynamics of a protein and the relative population distribution among its various structural states might be influenced by external conditions, most interestingly through the binding of ligands, small and large. The effect of ligand binding on the conformation and activity of a protein may occur far away from the binding site. This phenomenon was termed allostery by Monod and coworkers (10, 11), and remains a topic of active research (12, 13). While early conformation-based allostery models emphasized the thermodynamic states of the transforming proteins, it has been more recently proposed that allostery can also involve changes in conformational dynamics (14, 15). There has been much interest in defining the potential role of dynamics in allosteric transitions (16-19).

How fast are protein motions? The answer to this question is surprisingly complex, as conformational dynamics involve multiple time scales and amplitudes. Local structural changes, such as bond vibrations and transitions between side chain rotamers, occur on the femtosecond to nanosecond time scale. Motions of secondary structure elements, such as loops and helices, are slower, and even slower are the movements of tertiary and quaternary structure elements (domains and subunits, respectively) (9, 20, 21). In this Perspective we would like to focus on these large-scale conformational changes. As will be seen below, some theoretical estimates suggest that they can in fact be quite fast, taking only microseconds or even shorter times to complete. Experimental methods that can measure and trace large-scale (as opposed to local) motions on very short times have only appeared in recent years. We will introduce these methods, with some emphasis on single-molecule FRET spectroscopy, and provide several examples from our work and others' to demonstrate how fast motions can be probed, how they might be coupled to slower protein functional cycles, and, more generally, what we can learn from them on protein machine function.

**How fast can large-scale functional motions be?**

Consider a simple model of two protein domains connected with a flexible linker (Figure 1A). The relative motion of the two domains to close the gap between them is actually an abundant conformational change that is found in multiple proteins and is termed domain closure, or hinge-bending motion. We can model this relative motion using a simple Langevin equation that takes into account the spring potential (with the spring constant $\kappa$) and the viscous drag ($f$) by the aqueous solution (Figure 1B) (22):

(1) $f \frac{dx}{dt} + \kappa x = F$

where $x$ is the coordinate of motion and $F$ is a random force acting on the moving element. Analysis shows that the relaxation time characterizing the motion, $\tau$, is given by

(1) $\tau = \frac{f}{\kappa}$

Inserting a value for the viscous drag on a 100 kDa protein, ~60 pN·s/m, and a value of 5 pN/nm for $\kappa$ (22), one obtains a value of 12 ns, surprisingly short! In fact, a nanosecond relaxation time was already predicted in 1976 by McCammon et al. (23), based on their calculation of the force constant for the hinge bending motion of lysozyme.

Both the Langevin-based estimate and the calculation of McCammon et al. involve barrierless motion (Figure 1C). More recent simulations of domain closure (e.g. (24)) indicate a free-energy barrier for the motion (Figure 1C), which would automatically slow down the relaxation time. To estimate the contribution of a free-energy barrier, we can start by assuming that such a barrier is caused by the need to form or break one or more salt bridges connecting the two domains (a notion that goes back to Max Pertuz (25)). Surprisingly, there are only a few estimates for the breaking time of a salt bridge, based on computer simulations (26-28). Gruia et al. used a molecular dynamics simulation (27), to suggest a breaking time of ~200 ns for a salt bridge essential for the unfolding of staphylococcal nuclease. It is likely that this time will increase if more than a single salt bridge is involved in a conformational change. Indeed, Hyeon et al. used coarse-grained simulations to demonstrate that a multiple salt-bridge switching mechanism is responsible for conformational transitions in the molecular chaperone GroEL, with individual salt-bridge breaking and forming on the microsecond time scale (28). The overall conclusion is, therefore, that domain motions in proteins are intrinsically very fast, whereas slower domain motions are likely frustrated due to intramolecular interactions.

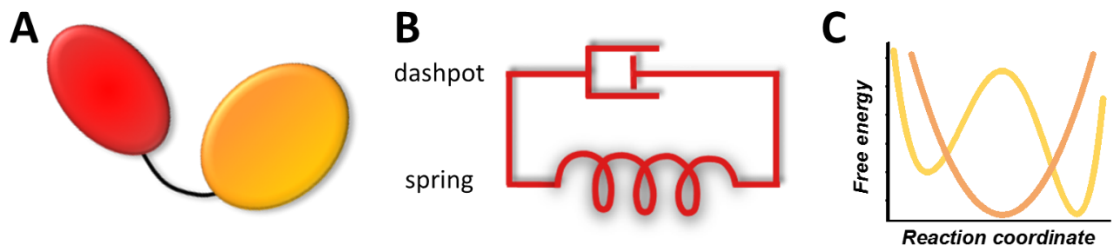

**Figure 1. Simple models for domain motions:** (A) Two protein domains connected with a flexible linker. (B) The relative motion of the two domains in A can be modeled using a spring coupled to a dashpot that represents the viscous drag on the moving parts. (C) While the model in B leads to a simple harmonic potential (orange), it is likely that the free-energy potential for realistic domain motions involves a barrier (yellow), perhaps due to breaking and forming salt bridges.

What about experimental estimates? Eaton and coworkers studied the propagation of signals between subunits in hemoglobin, based on optical spectroscopy of ligand rebinding to the heme moieties following photodissociation (29). They concluded that communication between subunits appears on the time range of 1-10 µs. Chakrapani and Auerbach used single-channel recordings of neuromuscular acetylcholine receptors to measure opening and closing rates (30). Analysis of multiple mutants allowed them to offer an upper limit to the channel-opening rate constant of ~0.9 µs$^{-1}$.

To summarize, there is a good reason to believe, based both on theoretical work and somewhat indirect experimental observations, that large-scale conformational changes in proteins should take place on the nanosecond-microsecond time scale. This immediately raises two questions:

1. Can we directly probe these motions experimentally and determine their rates?
2. How are such fast motions commensurate with the typically much slower function of many proteins?

We will start by answering the first question and showing how recent years have led to significant progress in our ability to probe fast motions in proteins. We will then discuss some examples from the literature and from our work, and comment, where possible, on potential answers to the second question.

**Methods for studying fast large-scale motions in biomacromolecules**

Among the rather small number of methods that can study protein tertiary structure dynamics, a special place is occupied by formidable nuclear magnetic resonance (NMR) spectroscopic techniques. Multiple NMR experiments have been developed to probe conformational changes on multiple time scales and with atomic resolution. These methods are based on a range of different experiments that are sensitive to a broad spectrum of relaxation times (31). A detailed introduction to the most useful methods for studying large-scale motion in proteins can be found in recent reviews (32, 33). Some of these techniques are suitable for situations where one of the probed species is 'invisible', i.e. significantly less populated than the other(s), but may detect kinetics down to the microsecond time scale. They include relaxation dispersion experiments, which quantifies changes in linewidth due to exchange between species, and paramagnetic relaxation enhancement measurements, in which the distance of a paramagnetic probe to protons is probed. Other methods, based on magnetization exchange or chemical exchange saturation transfer, are useful in situations where the populations of all detected states are substantial, but only probe slower time scales.

These powerful NMR techniques can provide a wealth of information on the populations of protein conformers and their exchange rates. There are additional spectroscopic methods that can probe large-scale dynamics within a protein molecule. For example, neutron spin echo spectroscopy measures an effective diffusion coefficient for internal motions within a protein, $D_{eff}(Q)$, as a function of the scattering vector $Q$ (37). Selective deuteration enables the measurement of motions of specific domains of a protein. Comparison to normal-mode analysis or molecular dynamics simulations can yield specific insights about relative motions of tertiary structure elements. While the above methods extract information on protein dynamics from equilibrium motions, time-resolved experimental techniques directly probe motions in the time domain by pushing the system under study out of equilibrium, using e.g. a laser-induced temperature jump or rapid mixing (38).

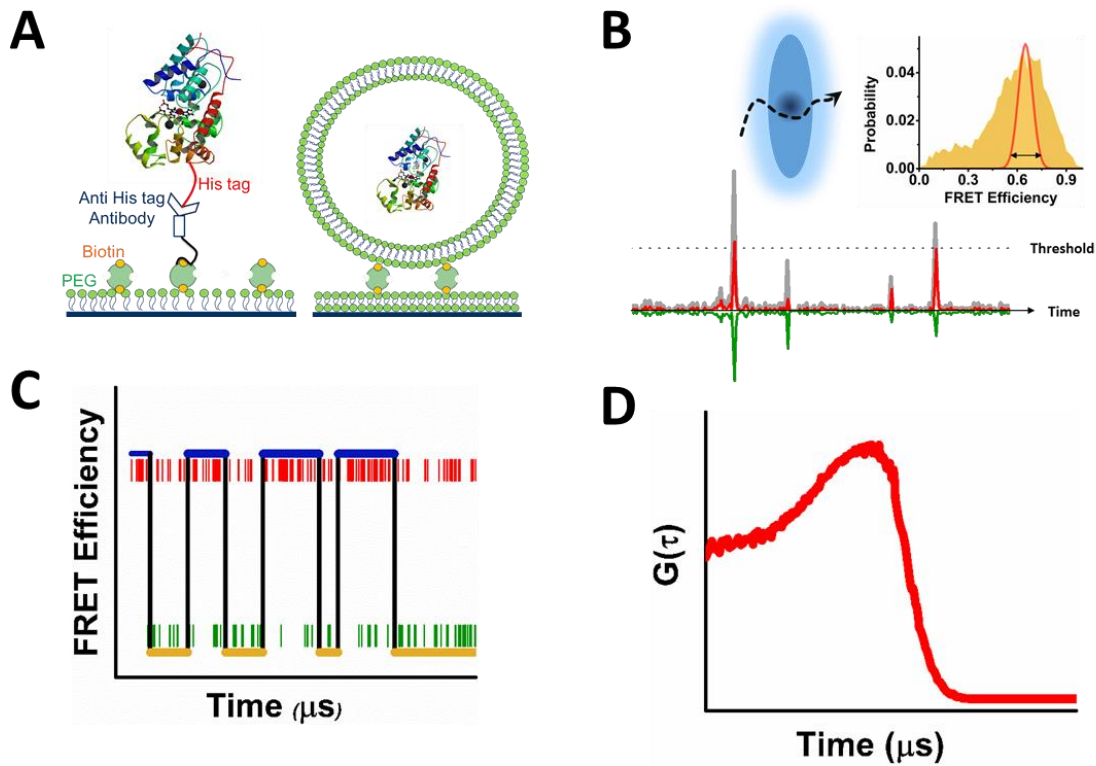

**Figure 2. Probing proteins with single-molecule fluorescence methodology:** (A) Slow conformational dynamics can be studied on surface-immobilized molecules. Two different methods to tether a single molecule to a surface are shown here, based on direct tethering or vesicle encapsulation. (B) Fast conformational dynamics can be studied on freely diffusion molecules. Top: cartoon of a laser focal volume with a molecule passing through. Bottom: fluorescence bursts emanating from excited molecules. Inset; FRET efficiency histogram (orange) that is broader than shot noise (white), suggesting fast dynamics. (C) Photon-by-photon single-molecule trajectory, with donor and acceptor in green and red, respectively. Blue and orange lines are state assignments from detailed statistical analysis. (D) Cross-correlation function of donor and acceptor fluorescence. The initial increase in the signal indicates conformational dynamics on the microsecond timescale. Adapted with permission from Ref [41].

Single-molecule experimental techniques complement ensemble methods, as they can probe real-time motions under equilibrium conditions. Single-molecule techniques characterize precisely conformational heterogeneity, observe conformational changes between multiple protein conformers as they occur and define the exact time constants typifying these dynamics. Single-molecule spectroscopy has therefore become critical to our understanding of molecular function, motion and dynamics. One group of powerful single-molecule spectroscopic tools that have been applied extensively to protein machines involves force spectroscopy (39), and includes optical and magnetic tweezers, as well as atomic force microscopy (40-42).

Here we focus on a second group of single-molecule methods, based on fluorescence, which are particularly useful for probing internal conformational changes of proteins with minimal intervention. Of the many single-molecule fluorescence spectroscopies introduced over the year, single-molecule fluorescence resonance energy transfer (smFRET) spectroscopy is particularly useful, as it can directly measure intramolecular distances within proteins, and in addition can also characterize the times and amplitudes of their modulation during function (43). smFRET experiments typically rely on excitation energy transfer between two fluorescent dyes attached to a protein, whose interaction depends strongly on their relative distance (44). In particular, in a typical smFRET experiment the signals from both a donor fluorophore and an acceptor fluorophore are measured. Any change in the conformation of the biomolecule to which these two dyes are attached leads to a change in the proportion of photons emitted by the donor vs. the acceptor.

A variety of smFRET experiments have been introduced over the years, covering multiple time scales from nanoseconds to seconds (45). For relatively slow time conformational dynamics (10 ms –s), smFRET experiments are conducted on molecules immobilized on a surface (Figure 2A). To probe fast conformational dynamics (ns - ms), smFRET spectroscopy can be more readily performed on molecules diffusing in solution (Figure 2B). In the context of this Perspective, focusing on fast conformational dynamics, we are interested in the latter type of experiments. Conformational changes can be characterized based on studies on diffusing molecules by either directly analyzing single-molecule trajectories (46) (Figure 2C) or performing fluorescence correlation spectroscopy (FCS) and calculating correlation functions from the signals (47) (Figure 2D).

**Ultrafast large-scale motions in proteins and their functional role**

The methods discussed in the previous section have been used extensively for *operando* observation of proteins, especially protein machines. We will introduce here a few notable recent examples in which very fast domain motions, on the sub-millisecond and microsecond time scale, have been detected.

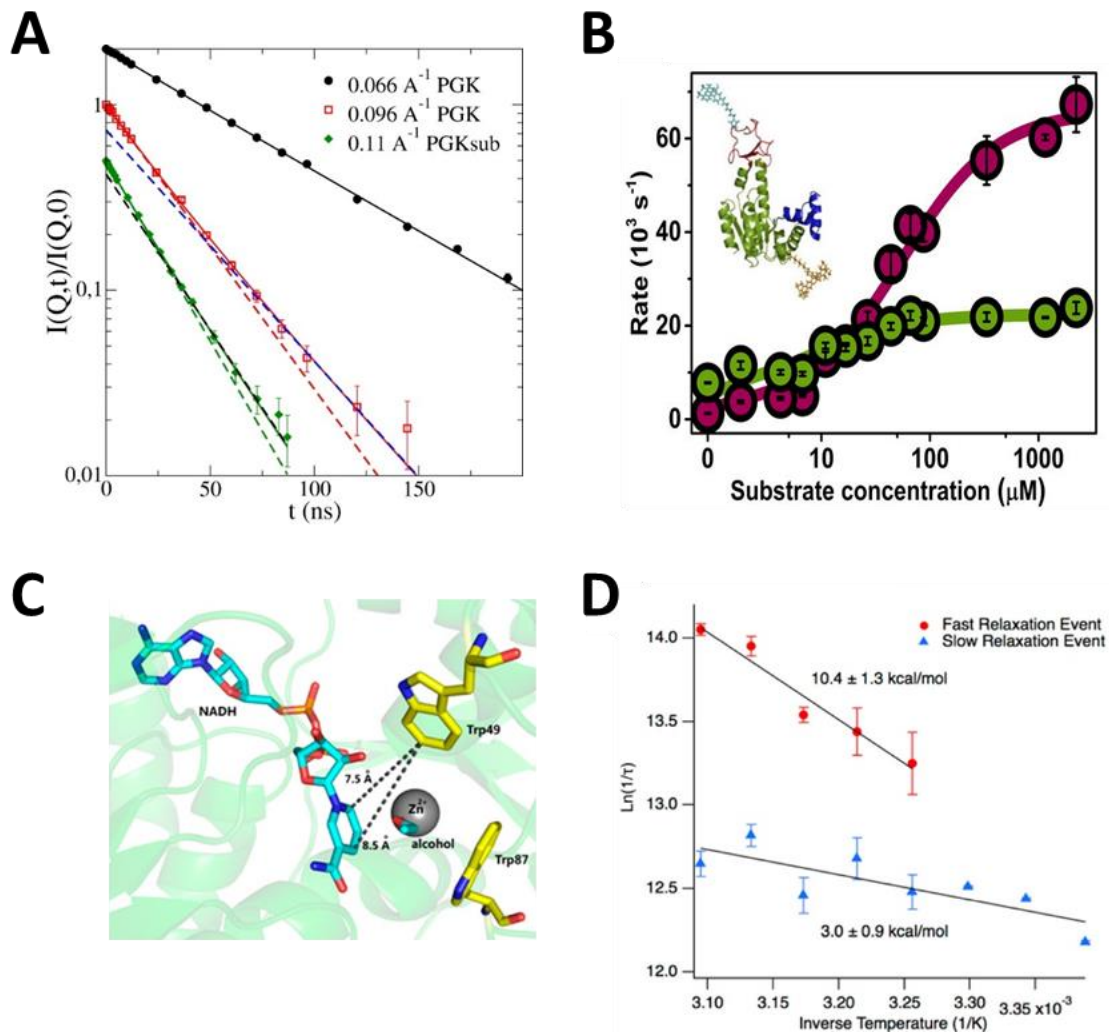

**Figure 3. Ultrafast domain motions in proteins and relation to function I:** (A) Phosphoglycerate kinase. Neutron spin echo intensity curves at different Q values. The curves at higher Q values do not fit a single-exponential line, which would arise from a simple overall diffusive motion of the proteins. Analysis exposes internal dynamics on a time scale of 50 ns. Reprinted from ref [47] with permission. (B) Adenylate kinase. Domain closing and opening rates (cherry and green, respectively) as a function of substrate concentration. Inset: structure of the protein with the labels attached. Reprinted from ref [19] with permission. (C-D) Alcohol dehydrogenase. FRET from tryptophans close to the active site to the NADH cofactor (as seen in the structure in C) is probed as a function of temperature. A fast relaxation component (D, red points) appears only at high temperatures. Reprinted from ref [51] with permission.

Phosphoglycerate kinase (PGK), is a glycolytic enzyme that catalyzes the reaction 1,3-bisphosphoglycerate + ADP ⇌ glycerate 3-phosphate + ATP, and has been shown to close the cleft between its two domains in order to bring the two substrates together for the chemical reaction. This dramatic conformational change has been termed a hinge-bending motion, as it involves a rigid-body rotation around a helix that connects the two domains (48). Neutron spin-echo spectroscopy was applied to probe the dynamics of PGK (49). Large-scale motions on a time scale of ~50 ns were identified (Figure 3A), and based on normal mode analysis, and later molecular dynamics simulations (50), it was concluded that these dynamics indeed involve a hinge-bending motion. A more recent molecular dynamics simulation suggested that the hinge-bending motion of PGK is in fact self-similar on timescales from $10^{-12}$ to $10^{-5}$ s (51). The fast domain closure dynamics in PGK are surprising, as the catalytic turnover rate of the enzyme is only 350 $s^{-1}$.

A similar situation was discovered in an experiment on a different protein whose reaction involves a domain closure step, adenylate kinase (AK) (19). AK is key to the maintenance of ATP levels in cells, catalyzing the reaction ATP+AMP ⇌ 2ADP. Domain closure is necessary in order to bring the two substrates together for the reaction. It has been suggested that the domain opening reaction is rate limiting for the overall enzymatic cycle of AK (52). Aviram et al. (19) measured the domain closure dynamics of freely diffusing AK molecules using smFRET spectroscopy. Photon-by-photon analysis of the data showed that, in the presence of substrates, AK's domain closure is in fact two orders of magnitude faster than its catalytic rate. It was found that ATP binding shortens significantly the domain closing and opening times, making them as short as 15 and 45 μs, respectively (Figure 3B).

The large mismatch between the time scale of conformational dynamics and the time scale of the enzymatic reactions of PGK and AK is surprising. A possible explanation comes from the realization that when an enzyme has to bind two or more substrates, ligand-binding disorder could prevent it from arriving at the right substrate configuration for the reaction. This implies that the substrates need to reorganize within the active site, which might be difficult to achieve in the closed conformation of the enzyme. Multiple domain opening and closing cycles should eventually bring the protein and its substrates to the conformation most conducive for the chemical reaction (19). Thus, the ultrafast domain closure dynamics in PGK and AK may serve to facilitate proper orientation of their substrates for the catalytic step. Interestingly, this conclusion is supported by findings of Klinman and coworkers on a different enzyme, thermophilic alcohol dehydrogenase (53). These authors performed temperature-jump ensemble measurements to probe time-dependent FRET between tryptophans close to the active site of the enzyme and the cofactor NADH (Figure 3C-D). They demonstrated a microsecond relaxation process that was activated above 30° C. The authors attributed this process to motions that, while being much faster than catalytic turnover, are able to facilitate the search for the appropriate configuration at the active site.

Ultrafast large-scale conformational changes were also detected in membrane proteins. For example, Shi et al. studied the dynamics of the rhomboid protease GlpG using solid-state NMR (54). Transmembrane helix 5 of GlpG serves as a gate and bends away from the rest of the protein to expose the catalytic dyad residues to the membrane environment. $R_{1\rho}$ RD measurements of this helix demonstrated opening and closing transitions on a time scale of 40 µs. Submillisecond conformational dynamics were discovered by Olofsson et al. in a study of the metabotropic glutamate receptor (mGluR), a G-protein-coupled receptor essential for synaptic activity (55). FRET-FCS experiments showed that the ligand binding domain homodimer of mGluR shuttles between two conformations, resting and active (Figure 4A-B). Surprisingly, the exchange between the two states was shown to be ultrafast, taking no more than 100 µs. Interestingly, the binding of ligands did not stabilize the protein in a single, active conformation as might have been expected. Rather, the population ratio between the resting and active states was continuously tuned. Barth et al. also used similar FCS techniques to demonstrate microsecond inter-domain dynamics in cohesion modules of the cellulosome (56).

Another example of fast dynamics on a time scale that is much faster than function is provided by ClpB, a AAA+ machine responsible for protein disaggregation in bacteria (57). Each of the six subunits of ClpB contains a regulatory switch, the middle domain (M domain) (Figure 4C). Mazal et al. (58) used smFRET to reveal transitions between the inactive and active states of the M domain on a time scale of ~150 µs (Figure 4D). M-domain motion is in fact much faster than the activity of the machine, suggesting that the ratio of the inactive and active conformations, rather than the population of one of them, serves to tune disaggregation. Indeed, factors that change this ratio, such as the concentration of the co-chaperone DnaK or nucleotides, were found to also modify the rate of disaggregation (Figure 4E). The mechanism of tunable allosteric switching reported by Olofsson et al. (55) and by Mazal et al. (58) likely involves a low-energy barrier between the active and inactive states. This may be a general way to modulate machine activity through analog (continuous) rather than digital (two-state) switching.

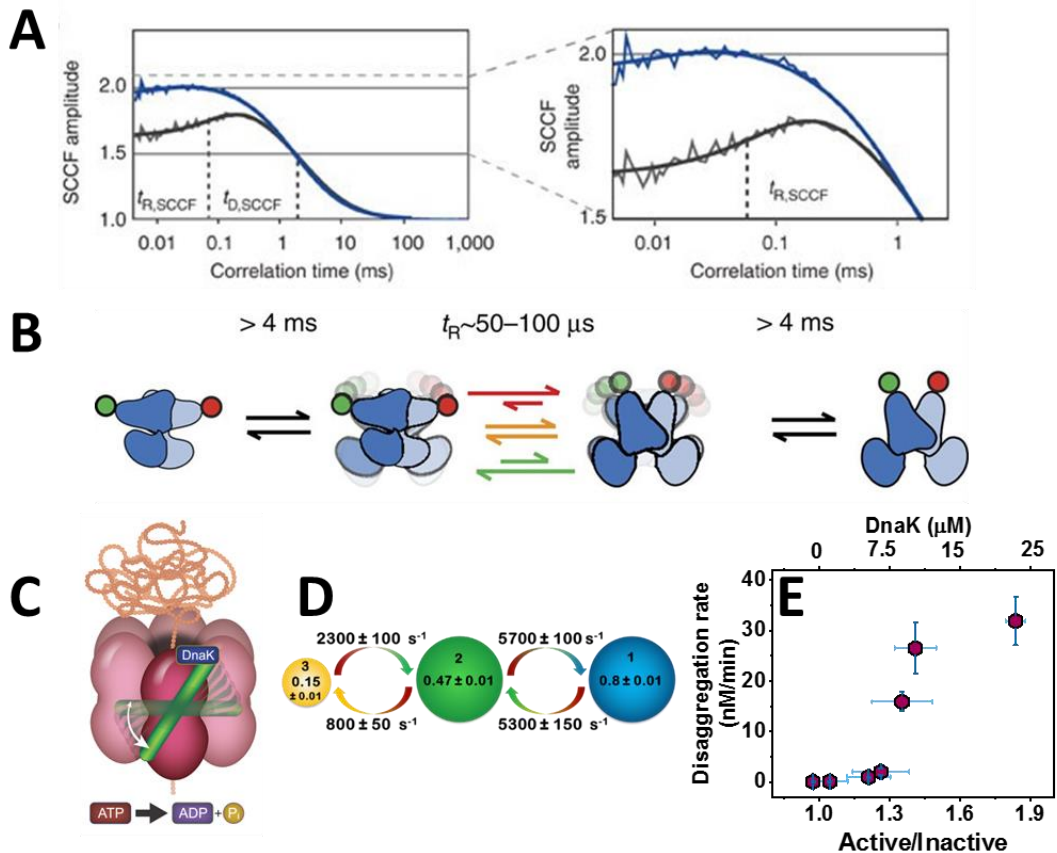

**Figure 4. Ultrafast domain motions in proteins and relation to function II:** (A-B) Dynamics of the G-protein-coupled receptor mGluR, reprinted from ref. [53] with permission. (A) Filtered fluorescence cross-correlation curves, indicating microsecond conformational dynamics in the wild type (grey) but not in the constitutively active mutant (blue). (B) kinetic model of mGluR dynamics. (C-E) The disaggregation machine ClpB, reprinted from ref. [56] with permission. (C) Schematic of ClpB and its dynamic M domain (green), which toggles between three states on the microsecond timescale. (D) Kinetic scheme derived from analysis of smFRET data. (E) DnaK binding tunes the ratio between states 1 and 2 of the M domain and in turn disaggregation activity.

**Conclusions and prospects**

The recent introduction of experimental methods that can measure large-scale dynamics of proteins on fast time scales has changed dramatically our ability to identify and characterize the motion of tertiary structural elements. In particular, the relative motion of domains and even subunits in proteins can now be directly studied. Both the extent (amplitude) of the motion and the typical timescale can be measured, even on the microsecond time scale. NMR and smFRET methods mentioned in this Perspective are particularly useful for probing fast dynamics, and may be seen as complementary to each other. Thus, NMR spectroscopies, particularly PRE-based, can characterize intermediate states in protein reactions with atomic resolution, but are not capable of directly measuring distance changes beyond 3.5 nm (36). smFRET spectroscopy, on the other hand, can only provide coarse structural information (although this situation can be improved by measuring multiple FRET pairs (59)), but is sensitive to larger distances and can be performed at low concentrations.

Based on the studies presented above, it is clear that tertiary structure motions in proteins can indeed be very fast. It is quite likely that such large-scale microsecond motions are much more ubiquitous than is currently assumed. Now, the overall functional cycles of proteins are often limited by chemical steps, such as ATP hydrolysis and related reactions, e.g. release of bound product molecules (60-62), and are therefore relatively slow. This is also true in the case of some of the examples given above: the enzymatic cycle of AK and PGK is much slower than the domain closure reactions; substrate threading in ClpB is much slower than the motion of the M domain; and the proteolytic activity of GlpG is much slower than motion of its helix 5. So how come that conformational dynamics are fast, while the functional cycles are slow?

There is no clear and general answer to this question. We pointed above to the possible role of fast dynamics in the search for the correct active-site configuration of substrates. Future experiments should establish how universal this conjecture is. Another role for fast conformational dynamics might be related to the mechanism of mechanical coupling in protein machines. Two limiting mechanisms discussed in the literature are the power stroke, in which the mechanical motion of a machine is fully coupled to nucleotide hydrolysis, and the Brownian ratchet, in which diffusive motions take place in between events of switching of the potential energy surface for motion between two states (63). Fast conformational dynamics might be relevant for the Brownian ratchet mechanism. For example, we have recently found (unpublished results) that the motion of ClpB's pore loops, i.e. the structural elements within the central pore of the machine that are responsible for substrate pulling, is very fast in comparison to substrate translocation time, supporting a Brownian ratchet mechanism for this machine.

A second question of interest in this relation is what physically makes functional cycles so much slower than conformational dynamics. One answer has to do with ligand binding and particularly ligand release. For example, the release of ADP following ATP hydrolysis is often rate limiting. Indeed, we have seen some time ago that the release of ADP from a large protein machine, GroEL, can be slow enough to prevent the protein from reaching equilibrium between its two main conformational states, T and R (64).

Experimental methodology is developing all the time. It is therefore reasonable to expect that the measurements of microsecond motions by equilibrium and non-equilibrium methods alike will become simpler and more robust in the not-to-far future. Improvements in experimental technique and data analysis should make studies of very fast protein dynamics simpler and therefore more abundant. We believe that a particularly thrilling future direction will be the application of these methods to study not only domain motions but also relative subunit motions in large, multi-subunit proteins. The connection of fast motions to the mechanisms of action of proteins is an endeavor that will likely require a combination of experimental results with theoretical/simulation work. This is thus an exciting frontier of protein dynamics that is expected to rapidly move forward in the coming years.

**Acknowledgements**

We would like to thank Drs. Marija Iljina and Amnon Horovitz for a critical reading of the manuscript. G.H. is the incumbent of the Hilda Pomeraniec Memorial Professorial Chair. This work was supported by a grant from the European Research Council (ERC) under the European Union's Horizon 2020 research and innovation programme (grant agreement No. 742637), and a grant from the Israel Science Foundation (No. 1250/19).

**Data Availability Statement**

Data sharing is not applicable to this article as no new data were created or analyzed in this study.